# High-resolution myelin-water fraction and quantitative relaxation mapping using 3D ViSTa-MR fingerprinting.

Congyu Liao[1,2], Xiaozhi Cao[1,2]*, Siddharth Srinivasan Iyer[1,3], Sophie Schauman[1,2], Zihan Zhou[1,4], Xiaoqian Yan[5,6], Quan Chen[1,2], Zhitao Li[1], Nan Wang[1,2], Ting Gong[7], Zhe Wu[8], Hongjian He[4,9], Jianhui Zhong[4,10], Yang Yang[11], Adam Kerr[2,12], Kalanit Grill-Spector[5] and Kawin Setsompop[1,2]

[1] Department of Radiology, Stanford University, Stanford, CA, USA

[2] Department of Electrical Engineering, Stanford University, Stanford, CA, USA

[3] Department of Electrical Engineering and Computer Science, Massachusetts Institute of Technology, Cambridge, MA, USA

[4] Center for Brain Imaging Science and Technology, College of Biomedical Engineering & Instrument Science, Zhejiang University, Hangzhou, Zhejiang, China

[5] Department of Psychology, Stanford University, Stanford, CA, USA

[6] Institute of Science and Technology for Brain-Inspired Intelligence, Fudan University, Shanghai, China

[7] Athinoula A. Martinos Center for Biomedical Imaging, Massachusetts General Hospital and Harvard Medical School, Charlestown, MA, USA

[8] Techna Institute, University Health Network, Toronto, ON, Canada

[9] School of Physics, Zhejiang University, Hangzhou, Zhejiang, China

[10] Department of Imaging Sciences, University of Rochester, Rochester, NY, USA

[11] Department of Radiology and Biomedical Imaging, University of California, San Francisco, CA, USA

[12] Stanford Center for Cognitive and Neurobiological Imaging, Stanford University, Stanford, CA, USA

* Corresponding author:

Xiaozhi Cao, PhD, xiaozhic@stanford.edu, Room 301, Packard Electrical Engineering Building, 350 Jane Stanford Way, Stanford, CA, 94305

# Abstract

**Purpose:** This study aims to develop a high-resolution whole-brain multi-parametric quantitative MRI approach for simultaneous mapping of myelin-water fraction (MWF), $T_1$, $T_2$, and proton-density (PD), all within a clinically feasible scan time.

**Methods:** We developed 3D ViSTa-MRF, which combined <u>Vi</u>sualization of <u>S</u>hort <u>T</u>ransverse rel<u>a</u>xation time component (ViSTa) technique with MR Fingerprinting (MRF), to achieve high-fidelity whole-brain MWF and $T_1/T_2/PD$ mapping on a clinical 3T scanner. To achieve fast acquisition and memory-efficient reconstruction, the ViSTa-MRF sequence leverages an optimized 3D tiny-golden-angle-shuffling spiral-projection acquisition and joint spatial-temporal subspace reconstruction with optimized preconditioning algorithm. With the proposed ViSTa-MRF approach, high-fidelity direct MWF mapping was achieved without a need for multi-compartment fitting that could introduce bias and/or noise from additional assumptions or priors.

**Results:** The in-vivo results demonstrate the effectiveness of the proposed acquisition and reconstruction framework to provide fast multi-parametric mapping with high SNR and good quality. The in-vivo results of 1mm- and 0.66mm-iso datasets indicate that the MWF values measured by the proposed method are consistent with standard ViSTa results that are 30x slower with lower SNR. Furthermore, we applied the proposed method to enable 5-minute whole-brain 1mm-iso assessment of MWF and $T_1/T_2/PD$ mappings for infant brain development and for post-mortem brain samples.

**Conclusions:** In this work, we have developed a 3D ViSTa-MRF technique that enables the acquisition of whole-brain MWF, quantitative $T_1$, $T_2$, and PD maps at 1mm and 0.66mm isotropic resolution in 5 and 15 minutes, respectively. This advancement allows for quantitative investigations of myelination changes in the brain.

# Introduction

Myelination is increasingly recognized as an important dynamic biomarker of brain development, aging, and various neurological conditions, including but not limited to multiple sclerosis, leukodystrophies, and neurodegenerative disorders (1–3). These conditions can lead to alterations in the biophysical properties observed in MRI scans (4–6). To quantitatively assess myelin content of brain, various quantitative MRI techniques (2,5,7–10) have been proposed based on different tissue properties, such as longitudinal relaxation time ($T_1$) (6,11), and transversal relaxation time ($T_2$) (12,13), $T_2^*$ (14,15), $T_1$-weighted/$T_2$-weighted (16), diffusion (17), and magnetization transfer (18–20). For instance, $T_1$ or $R_1$ (reciprocal of $T_1$) has been used to predict the myelination process in cortex (4,6,21), as the decrease in $T_1$ correlates with more myelination. Compared to $T_1$ which can be impacted by contributions from both myelin content and other confounds, myelin water fraction (MWF) (2,14,22) that probes short $T_1$ and $T_2$ signal contributions of water molecules trapped within the myelin sheaths has been shown to be more specific for predicting myelination changes and characterizing the myelin content in the brain (2,23).

To estimate MWF, conventional MWF mapping relies on a multi-echo spin-echo or gradient-echo sequence (22) and multi-compartment fitting of the exponential decay signal to extract the shorter relaxation time of myelin water (14,22,24). However, the acquisition time of the conventional method is long (e.g., 1 minute per 1 mm slice (22)), and the fitting process is ill-conditioned and susceptible to noise. To improve MWF mapping, <u>Vi</u>sualization of <u>S</u>hort <u>T</u>ransverse rel<u>a</u>xation time component (ViSTa) technique (25) was proposed for direct visualization of myelin water signal. This technique employs a specifically configured double inversion-recovery sequence that suppresses the long $T_1$ component while preserving the signals from the short $T_1$ components of myelin water. This allows for direct and precise imaging of myelin water, enabling accurate assessment of myelin content without fitting. However, ViSTa faces challenges such as decreased SNR due to signal suppression and a long acquisition time (40 seconds per slice with 1mm$^2$ even with 9x parallel imaging acceleration achieved via advanced wave-CAIPI techniques (26)).

Magnetic resonance fingerprinting (MRF) (27) is a rapid quantitative imaging technique that simultaneously estimates multiple tissue parameters and has garnered significant interest as a diagnostic tool in various diseases (28–31). This technique was initially proposed using a 2D acquisition. Since then, numerous studies have focused on advancing MRF to achieve shorter scan times, higher resolutions, improved accuracy and reduced variability. To enable fast high-resolution MRF for whole-brain quantitative imaging, 3D stack-of-spiral (32,33) and spiral-projection-imaging trajectory (34) have been developed. These advancements allow for whole-brain 3D MRF at 1 mm isotropic resolution in ~6 minutes. On the reconstruction side, various methods such as parallel imaging (32,35), low rank/subspace model (36–39) and deep learning methods (40–42) have been incorporated into MRF reconstruction to enhance image quality. Recently, MWF mapping has been conducted using modified MRF sequences that aims to achieve better signal separability between myelin water and other tissue types (5,43). However, the extraction of MWF still relies on multi-compartment fitting, which can pose challenges in accurately separating different components, particularly in highly undersampled MRF data with low signal-to-noise ratio (SNR). The multicompartment fitting is ill-posed and typically requires additional assumptions and/or priors to obtain good results, which could create bias or artifacts. This is an open area of research, where a number of innovative reconstruction algorithms (44–47) are being developed to tackle this issue.

In this work, we have developed a novel 3D ViSTa-MRF acquisition and reconstruction framework that integrates the ViSTa technique into MRF. This approach achieves a remarkable acceleration of MWF mapping by 30x compared to the gold standard ViSTa approach (1.3 seconds per slice with 1mm$^3$ resolution) while also enabling better SNR and simultaneous estimation of $T_1$, $T_2$, and PD. With the double inversion prep in 3D ViSTa-MRF, we can directly visualize MWF image once the time-series data is reconstructed, without the need for multicompartment modeling. We demonstrate that the proposed method achieves high-fidelity whole-brain

MWF/$T_1$/$T_2$/PD maps at 1mm and 0.66 mm-isotropic resolution in 5 minutes and 15.2 minutes, respectively. Furthermore, we propose a 5-minute whole-brain 1mm-iso ViSTa-MRF protocol to quantitively investigate brain development in early childhood. This work is an extension of our earlier work, which was reported as conference abstract and oral presentation in the Annual Meetings of International Society of Magnetic Resonance in Medicine (ISMRM) 2022 (48).

## Methods

### ViSTa-MRF sequence

Figure 1(A) shows the sequence diagram of the ViSTa-MRF acquisition, where each acquisition group consists of multiple ViSTa-preparation blocks and one MRF-block. A water-exciting rectangular (WE-Rect) hard pulse (49) was employed for signal excitation, where the RF duration was set to 2.38ms at 3T so that the first zero-crossing of its sinc-shaped frequency response is at the main fat-frequency (440 Hz). In each ViSTa-block, specifically configured double-inversion-recovery pulses were applied ($TI_1$=560ms, $TI_2$=220ms), and the first subsequent signal time-point was referred to as the "ViSTa signal". Twenty consecutive time points were acquired within each ViSTa-block to facilitate joint spatial-temporal subspace reconstruction. Through extended-phase-graph (EPG) (50) simulation, Figure 1(B) shows that the myelin-water signal was preserved in the ViSTa signal, while the white-matter (WM), gray-matter (GM) and Cerebrospinal fluid (CSF) were suppressed, which enabled direct myelin-water imaging. At the end of the ViSTa-block, a BIR-4 90° saturation-pulse with a spoiler gradient was applied to suppress flow-in CSF and vessel signals. A waiting time (TD) of 380ms was selected to achieve a steady-state longitudinal magnetization of the short-$T_1$ signal for the next ViSTa preparation. To enhance the encoding of the short-T1 signal, the sequence repeated the ViSTa-block eight times, followed by an MRF block, resulting in a total acquisition time of 19 seconds for each acquisition group. Increasing the number of ViSTa blocks yielded more ViSTa signal encodings but extended acquisition time. To establish the optimal number of required ViSTa blocks, we undertook empirical tests using varying quantities of ViSTa blocks and assessed the reconstructed ViSTa image quality to guarantee its likeness to the standard ViSTa sequence. Through our experiments, we identified that employing 8 ViSTa blocks struck the ideal balance between ViSTa signal quality and acquisition duration. In the last ViSTa-block, the saturation pulse and TD time were omitted to ensure a smooth signal transition between the ViSTa block and the MRF block. This step was taken to standardize the signal across all eight ViSTa blocks and to prepare for obtaining the ViSTa signal, which was not required in the last block. After the ViSTa-blocks, 500-time-point FISP-MRF (51) block were acquired. Unlike conventional MRF acquisition, where the inversion-recovery pulse was placed at the beginning of the MRF block. In this approach, we introduced a 1-second rest time before applying the inversion-recovery pulse at the 200th time-point of the MRF block. This design allows for the recovery of longitudinal magnetization after ViSTa preparations. Between the acquisition groups, a BIR-4 90°-saturation-pulse with a TD of 380ms was used to suppress flow-in CSF and vessel signals and achieved steady-state longitudinal-magnetization of short-$T_1$ signal.

To improve the SNR and the estimation accuracy of myelin water ($T_1/T_2$ = 120/20ms), WM ($T_1/T_2$ = 750/60ms), and GM ($T_1/T_2$ = 1300/75ms), we employed the Cramér–Rao lower bound (CRLB) of $T_1/T_2$ values to optimize the flip angle (FA) train in the ViSTa-MRF sequence (52,53). Figure 1(C) shows the ViSTa-MRF signal-curves with good signal-separability between different tissue-types. To achieve efficient sampling in 3D k-space, the ViSTa-MRF sequence employed the optimized 3D tiny-golden-angle-shuffling (TGAS) spiral-projection acquisition (39) with 220×220×220 mm$^3$ whole-brain coverage for incoherent undersampling. Different numbers of acquisition groups were employed for different resolution cases, where the 3D TGAS were designed to rotate around three axes, as shown in Figure 2(A).

**Synergistic subspace reconstruction**

We proposed a memory-efficient fast reconstruction that leverages spatial-temporal subspace reconstruction (36–38,54) with optimized k-space preconditioning (55). The ViSTa-MRF dictionary, accounting for $B_1^+$ variations ($B_1^+$ range [0.70:0.05:1.20]), was generated using EPG, and the first 14 principal components were chosen as the temporal bases (Figure 2(A)). Compared to our previous study (39), where only 5 bases were selected in the subspace reconstruction for a conventional MRF sequence, in ViSTa-MRF, 14 bases were used in the present ViSTa-MRF study to better represent the myelin-water signal. This change was due to the sequence design of ViSTa-MRF, which introduced more signal variations through CRLB optimization. To determine the number of bases needed for the reconstruction, we applied two conditions: (i) Ensuring that the number of bases is sufficient to represent 99% of the signal in the dictionary. (ii) Conducting empirical tests with different numbers of bases and examining the quality of ViSTa signal to ensure that it resembles the results obtained from the standard ViSTa sequence. The ViSTa-MRF time-series was projected onto the subspace, resulting in 14 coefficient maps based on the selected temporal bases. The ViSTa-MRF time-series $\mathbf{x}$ is expressed as $\mathbf{x}=\mathbf{\Phi c}$, where $\mathbf{\Phi}$ are the temporal bases, $\mathbf{c}$ are the coefficient maps. Figure 2(B) illustrates the flowchart of the subspace reconstruction with locally low-rank constraints, which could be described as:

$$min_{\mathbf{c}} \|\mathbf{MFS\Phi c} - \mathbf{k}\|_2^2 + \lambda \|\mathbf{c}\|_*  \qquad [1]$$

where $\mathbf{S}$ contains coil sensitivities, $\mathbf{F}$ is the NUFFT operator, $\mathbf{M}$ is the undersampling-pattern, $\lambda$ is the regularization-parameters. We implemented a novel algorithm in SigPy(56) to solve Equation [1] that combined polynomial preconditioned FISTA reconstruction with Pipe-Menon density-compensation (55) and basis-balancing (57) to reduce artifacts and accelerate the subspace reconstruction. The off-line reconstruction package is available at https://github.com/SophieSchau/MRF_demo_ISMRM2022. With this reconstruction approach, the whole-brain 14-bases coefficient maps (e.g., $220_x \times 220_y \times 220_z \times 14_{bases}$ for 1mm-iso MRF data) can be efficiently reconstructed on a GPU with 24 GB VRAM in 45 minutes (~90s per iteration, 30 iterations with polynomial preconditioning). This provides a significant improvement compared to reconstruction performed using e.g., the popular BART software(58) which requires over 1TB of RAM and reconstruction time of over 4 hours on our Linux server for this same problem. This advancement allows for fast reconstruction of large spatial-temporal data, providing more efficient processing. As shown in Figure 2(B) and (C), the reconstructed coefficient maps (**c**) were then used to generate the time-series with voxel-by-voxel $B_1^+$ correction for estimating $T_1$/$T_2$/PD maps, while the quantitative MWF map was derived from the reconstructed first time-point ViSTa image I(ViSTa) and the PD map I(PD):

$$\mathrm{MWF} = \frac{\mathrm{I(ViSTa)}}{\mathrm{I(PD)} \cdot \mathrm{S(myelin\_water)}}, \qquad [2]$$

where S(myelin_water) is the $B_1^+$ corrected, EPG-simulated signal intensity from the dictionary using nominal $T_1$ and $T_2$ values of myelin-water ($T_1$/$T_2$ =120/20ms). The S(myelin_water) signal is normalized to '1', where the term '1' denotes the maximum signal tipped down by a 90-degree excitation pulse from a fully recovered Mz.

Using spatiotemporal subspace reconstruction, the entire time-series was jointly reconstructed, allowing us to reconstruct the first time-point image (ViSTa signal) and leverage the encoding and SNR-averaging from all other time points. The implementation of the ViSTa-MRF acquisition and joint subspace reconstruction enabled us to directly visualize myelin-water images once the time-series data were reconstructed. It is important to highlight that the myelin-water signal evolution throughout the MRF sequence (as shown in Figure 1(C)) is markedly different from that of WM and GM signals throughout the entire sequence. This unique behavior of the myelin-water signal, along with the subspace reconstruction, effectively utilized the signal and spatial encoding throughout the MRF sequence to differentiate the myelin-water signal from other tissue types and to create a high SNR first ViSTa timepoint data. This capability would not have been achievable with, for example, sliding window NUFFT reconstruction (59).

By utilizing the reconstructed quantitative $T_1$, $T_2$, and PD maps, we can synthesize multiple contrast-weighted images using Bloch equation that provide robust contrasts while significantly reducing scan time and improving motion-robustness during examinations.

**In-vivo acquisition and reconstruction**

We implemented 1.0 mm and 0.66 mm isotropic whole-brain ViSTa-MRF sequences on one 3T GE Premier scanner and one ultra-high-performance (UHP) scanner (GE Healthcare, Madison, WI, USA), as well as two 3T Siemens Prisma scanners and one Vida scanner (Siemens Healthineers, Erlangen, Germany). A total of twenty healthy adults (age: 23.4 $\pm$ 2.3 year-old) participated in the study, with approval from the institutional review board. Written informed consent was obtained from each participant. FOV: 220×220×220 mm$^3$, TR/TE=12/1.8ms with a 6.8ms spiral-out readout and a 1.2ms rewinder for both 1mm and 0.66mm cases. The maximum gradient strength was 40mT/m, and the maximum slew rate was 100T/m/s for 1mm resolution. For the 0.66mm resolution, the maximum gradient strength was 60mT/m, and the maximum slew rate was 160T/m/s. Sixteen and forty-eight acquisition-groups with eight ViSTa-blocks were acquired for 1-mm and 0.66-mm cases, respectively, to achieve sufficient spatiotemporal encoding. This resulted in scan times of 19s×16=5 minutes for the 1mm-iso and 19s×48=15.2 minutes for the 0.66mm-iso datasets. FOV-matched low-resolution (3.4mm×3.4mm×5.0mm) $B_0$ maps were obtained using multi-echo gradient-echo sequence. To mitigate $B_0$-induced image blurring from the spiral readout, a multi-frequency interpolation (MFI) technique (39,60) was implemented in subspace reconstruction with conjugate phase demodulation. To achieve robustness to $B_1^+$ inhomogeneity, $B_1^+$ variations were simulated into the dictionary and incorporated into the subspace reconstruction. Bloch-Siegert method was utilized to obtain FOV-matched low-resolution $B_1^+$ maps (3.4mm×3.4mm×5.0mm). These low-resolution $B_1^+$ maps were then linearly interpolated to match the matrix size of the high-resolution ViSTa-MRF results, as the $B_1^+$ maps are spatial smoothing. As the ViSTa-MRF dictionary included $B_1^+$ effects, the obtained $B_1^+$ maps were used to select a sub-dictionary for matching at each pixel, thus corrected for $B_1^+$ inhomogeneity related $T_1$ and $T_2$ bias in ViSTa-MRF (61). The quantitative maps with and without $B_1^+$ corrections were compared.

The 20 adult datasets were acquired from five scanners. To test the cross-scanner comparability, we selected the data from one Prisma scanner as the reference. To calculate the cross-scanner mean $T_1$, $T_2$, and MWF values, 32 representative WM and GM regions, along with 5 representative MWF regions were chosen. Using these mean values, the reproducibility coefficient (RPC) and Bland–Altman plots for $T_1$, $T_2$, and MWF were computed.

In order to assess the performance of the CRLB-optimized protocol, we acquired the ViSTa-MRF sequence using both the original FAs and the CRLB-optimized FAs. The original FAs and the CRLB-optimized FAs are depicted in Figure 3(A). To evaluate the fat artifacts, the WE-Rect pulse is compared with the normal non-selective Fermi pulse we used in our previous study. To validate the accuracy of myelin estimation of the proposed ViSTa-MRF method, the proposed method is compared with the standard 2D fully sampled ViSTa sequence with multi-shot spiral readout. The in-plane resolution of the ViSTa is 1mm, slice thickness 5mm. 48 spiral interleaves are used to fully sample one slice, results in the total acquisition time of 48× ($TI_1$ 560ms+$TI_2$ 220ms +TD 380ms) = 56s per slice. The acquisition was not accelerated with parallel imaging as the reconstructed image with full sampling was already at low SNR. This is much slower than ViSTa-MRF, as the proposed 1mm-iso ViSTa-MRF could acquire 220 slices in 5 minutes (1.3s per slice). To validate the accuracy of $T_1$ and $T_2$ estimation of the ViSTa-MRF method, the proposed method is compared with the standard 3D MRF sequence at 1mm isotropic resolution.

In addition to validating our approach on healthy adult volunteers, data were also acquired on two infants to quantitatively investigate infant brain development using MWF, $T_1$ and $T_2$ maps. A 5-minute whole-brain ViSTa-MRF and a $T_1$-MPRAGE (magnetization-prepared rapid gradient echo) with 1.0 mm-isotropic resolution were utilized to acquire data on a 4-month and a 12-month infant. The protocol parameters of the $T_1$-MPRAGE sequence for the baby scans are: TR/TE= 6.9/2.3ms, inversion time = 400ms, flip-angle=11°, image resolution 1×1×1mm$^3$, FOV= 220×220 ×220mm$^3$, total acquisition time: 6 minutes 20 seconds. The experiments were performed on a 3T GE UHP scanner with the approval of the institutional review board. Written informed consent was obtained from each participant's parents. Scans were scheduled approximately 1 hour after the infant's bedtime, for a 2-hour window to allow sufficient time for the infant to fall asleep and/or restart scans if the infant awakens. MR-compatible headphones for infants were used to ensure appropriate noise protection during the scans.

**Ex-vivo scans**

Additional data were also acquired on ex-vivo brain samples where long scan time is feasible to investigate capability of ViSTa-MRF for investigation of mesoscale quantitative tissue parameter mapping. To validate the image quality of the proposed method, a coronal slab from a 5-month-old post-mortem brain and a left occipital lobe sample from a 69-year-old post-mortem brain were acquired with ViSTa-MRF at 0.50mm-isotropic resolution: FOV:160×160×160mm$^3$, 180 acquisition-groups were acquired with total acquisition time of 19s×180=57minutes. For the ex-vivo scans, a lower acceleration rate than feasible was used to ensure high SNR.

**Results**

Figure 3(B) shows reconstructed 1mm-iso $T_2$ and MWF maps acquired from a healthy adult using the original FAs and the CRLB-optimized FAs. The red arrows indicate the CRLB-optimized results achieve higher SNR in MWF maps. The zoom-in figures demonstrate that the CRLB-optimized ViSTa images exhibit higher SNR and better visualization of detailed structures in the cerebellum than the ViSTa-MRF images with the original FAs.

Figure 4(A) shows a representative time-resolved 1mm-iso MRF-volume after subspace reconstruction using the original fermi-pulse and the WE-Rect pulse. As yellow arrows indicate, the fat artifacts are much mitigated using the WE-Rect pulse. Figure 4(B) shows the comparison between a fully sampled standard 2D-ViSTa sequence (56s/slice) and ViSTa-MRF acquisition (1.3s/slice) with subspace reconstruction, where the results are highly consistent, demonstrating the feasibility in leveraging the joint-spatiotemporal encoding information for highly accelerated ViSTa-MRF data. Figure 4(C) shows $T_1$, $T_2$ and MWF maps with and without $B_1^+$ correction as well as corresponding $B_1^+$ maps. With $B_1^+$ correction, the estimated $T_2$ and MWF maps are more uniform compared to the results without $B_1^+$ correction.

Figure S1 shows a representative slice of $T_1$ and $T_2$ maps estimated from standard MRF and ViSTa-MRF methods. The comparison between the 1mm ViSTa-MRF and the standard MRF sequences demonstrates that the quantitative $T_1$ and $T_2$ maps estimated from ViSTa-MRF are highly consistent with the standard MRF sequence.

Figure 5(A) shows the 5-minute whole-brain 1mm-iso $T_1$, $T_2$ and MWF maps in coronal views, where the MWF values for a healthy adult shown in Figure 5(B) from ViSTa-MRF across four representative WM-regions: genu corpus callosum, forceps minor, forceps major and corpus callosum splenium, are consistent with the literature results (25). The MWF comparison between literature values and our proposed ViSTa-MRF method is shown in Table S1. The region of interest (ROI) size is 5×5 for the four WM regions.

Figure 6 displays whole-brain 660μm $T_1$, $T_2$, PD, ViSTa, and MWF maps obtained within a 15-minute scan time. The zoom-in figures highlight the enhanced ability to visualize subtle brain structures, such as the caudate nucleus (red arrows in Figure 6). When compared to the 1mm results, the higher resolution of the 660μm dataset provides improved visualization of the periventricular space (red arrows in Figure S2).

Figure S3 shows the cross-scanner comparability of ViSTa-MRF data acquired from 5 scanners and the Bland–Altman plots for $T_1$, $T_2$, and MWF. The results demonstrate robust ViSTa-MRF results across different scanners.

Figure 7(A) shows the estimated 1mm-iso whole-brain $T_1$, $T_2$, and MWF maps of 4-months, 12-months babies and a reference 22-year-old adult. As shown in Figure 7(B), a custom-built tight-fitting 32-channel baby coil (62) is used to acquire datasets for improved SNR. The estimated $T_1$ and $T_2$ values of white-matter and gray-matter decrease while the estimated MWF of white-matter increases with brain development, indicating brain dynamic process of myelination, as shown in Figure 7(C).

By leveraging the fast acquisition of ViSTa-MRF, we were able to synthesize various contrast-weighted images across the whole-brain at high resolution from the 12-month-old infant data, including $T_1$-MPRAGE, $T_1$- and $T_2$-weighted, $T_2$-FLAIR (Fluid attenuated inversion recovery) and DIR (double inversion recovery) images, as shown in Figure 8(A). The parameters for these sequences were optimized to account for the typical $T_1$ and $T_2$ of various tissues in infant brain. This eliminates the need for time-consuming structural scans in infant studies and provides an alternative when motion artifacts compromise the quality of conventional contrast-weighted images. Figure 8(B) provides a visual comparison between an acquired $T_1$-MPRAGE image and a synthesized image. The acquired $T_1$-MPRAGE scan at 0.9mm, which requires a 12-minute acquisition, was adversely affected by subject motion, resulting in image blurring and compromised image quality, as highlighted by the red arrow in the zoom-in figure in Figure 8(B). In contrast, our fast ViSTa-MRF acquisition produced a synthesized $T_1$-MPRAGE image with improved quality and CNR, where the white and gray-matter contrast was maximized through synthetic sequence parameters selection using Bloch simulation.

Figure 9 shows the 0.50mm-iso ViSTa-MRF results of the post-mortem brain samples. Figure 9 (A) shows quantitative $T_1$, $T_2$, MWF and PD maps obtained from an ex-vivo 5-month infant brain. The zoom-in figure in Figure 9 (A) reveals decreased $T_1$, $T_2$, PD, and increased MWF values (indicated by red arrows) in the lines of Baillarger within cortical layers, reflecting the higher myelination level in the cortex. Figure 9(B) shows ViSTa-MRF results of the 69-year-old post-mortem brain sample. The decreased $T_1$ and PD and increased MWF values (indicated by black arrows) were detected in the lines of Gennari in V1 region, reflecting the high myelination in Layer IV of the cortex, which are consistent with the high-resolution $T_2$-weighted reference images. As red arrow indicated in Figure 9(B), the "dark dots" in MWF and increased $T_1$ and $T_2$ values imply the de-myelination in this region in the aging brain. These findings align with results from other studies (63–65).

**Discussion**

In this work, we developed a 3D ViSTa-MRF sequence with a CRLB-optimized FAs and a memory-efficient subspace reconstruction to achieve high-resolution MWF, $T_1$, $T_2$, and PD mapping in a single scan. Compared to the accurate yet time-consuming standard ViSTa sequence, the proposed fast ViSTa-MRF approach provides consistent MWF values at 30x faster scan time with higher SNR. We demonstrate that the proposed method achieves high-fidelity whole-brain MWF, $T_1$, $T_2$, and PD maps at 1mm- and 0.66mm-isotropic resolution on 3T clinical scanners in 5 minutes and 15.2 minutes, respectively. Furthermore, our preliminary results of the 5-minute infant scans demonstrate the feasibility in using this technology for investigating brain development in early childhood.

Previous studies (28,44–46) have demonstrated that MRF emerges as a promising multi-contrast acquisition strategy capable of estimating multi-compartment quantitative tissue parameters within a shorter duration. However, it has been recognized that conventional MRF techniques may have limited sensitivity to tissue compartments characterized by short $T_1$ values, such as the myelin water component in brain tissue (66). To address this limitation, several methods have been proposed to modify the MRF sequence. For example, the incorporation of ultra-short-TE acquisition (12) has been used to extract and differentiate the ultra-short $T_2$ component of pure myelin from the long $T_2$ component of myelin-water signals (67). Additionally, a multi-inversion preparation with short inversion times has been added to the MRF sequence to improve the sensitivity of the short $T_1$ myelin-water signal (5,43). These modifications aim to enhance the accuracy and specificity of MRF in quantifying myelin-related tissue property. However, these approaches still have limitations, as they either require multi-component fitting (e.g., non-negative least squares with joint sparse constraints (45,46)) or rely on strong assumptions of predefined compartments with fixed $T_1$ and $T_2$ values (5). These assumptions can be sensitive to noise or may introduce biases in MWF quantification. In contrast, our proposed ViSTa-MRF method with CRLB optimization demonstrates promising sensitivity to short $T_1$ components, eliminating the need for multi-compartment modeling or predefined dictionary fitting with fixed $T_1$ and $T_2$ values.

In standard ViSTa acquisition with a TR of ~1.3s, 0.8s is used for the inversion preps, 10ms for short readout due to the fast $T_2^*$ decay of myelin-water signal and 0.4s for recovery. This is highly inefficient. In this work, we incorporate ViSTa into MRF, to improve the speed and accuracy of MWF-mapping. The idea of ViSTa-MRF is to combine original ViSTa sequence with MRF, where the first time point provides pure myelin water signal (ViSTa signal) while the subsequent MRF time-point provide signal from multi-tissue compartment, with each compartment having a CRLB-optimized distinct/non-overlapping signal evolution. Using spatiotemporal subspace reconstruction, the whole time-series (19s per acquisition group) is jointly reconstructed to enable the reconstruction of the first time-point (ViSTa signal) image to leverage the encoding and SNR-averaging from all the other time points. With the implementation of the ViSTa-MRF acquisition and joint subspace reconstruction, we gain the ability to directly visualize myelin-water images once the time-series data is reconstructed. Additionally, we can achieve high-resolution whole-brain quantitative $T_1$, $T_2$, and PD maps, just like the original MRF method. Moreover, unlike the standard ViSTa method, which necessitates an additional gradient echo image to calculate final MWF maps, the ViSTa-MRF approach allows for the calculation of MWF using the MRF-estimated PD maps without the need for any additional scans.

The ViSTa signal has been shown to primarily consist of water components with short relaxation times, and research indicates that this signal predominantly originates from myelin water (25). However, it is important to consider other influencing factors such as cross relaxation, myelin water exchange, and magnetic transfer, which could potentially lead to an underestimation of MWF values (26,68). Due to these complexities, the MWF maps derived from ViSTa are also termed as apparent MWF maps (23). These maps continue to serve as indicators of myelin content and are employed for both intra- and inter-subject comparisons, as well as cross-sectional and longitudinal studies (69,70). In our ViSTa-MRF simulation, we simplified the model to a single pool without the magnetization transfer effect, which could lead to a potential underestimation of $T_1$ and $T_2$ values.

In this work, we employed the CRLB for the optimization of the ViSTa-MRF sequence. The optimized flip angle for the first time-point was determined to be 38°, which differs from the standard ViSTa sequence that uses a 90° excitation. This variation is attributed to the differences in the acquisition methods. While the standard ViSTa sequence utilizes a single readout to maximize the signal with 90° excitation, the ViSTa-MRF acquisition employs continuous readouts to leverage the myelin-water component and achieve SNR-averaging across all time-points with the subspace reconstruction. To strike a balance between high SNR in the first time-point and distinct signal evaluation of the myelin-water component in the continuous readouts, CRLB optimization was employed for the flip-angle train of ViSTa-MRF. Consequently, our in vivo comparison revealed improved SNR in the ViSTa image when compared to the non-optimized protocol. Furthermore, we also applied the ViSTa-MRF sequence for infant scans to quantitatively investigate infant brain development. Given the rapid changes in relaxation times during the development of infant brains, our future work will involve calculating the CRLB for age-specific $T_1$ and $T_2$ values of myelin-water, white matter, and gray matter. This optimization will enable us to tailor the infant scan protocol to different ages of infants, ensuring accurate and sensitive assessments of brain development.

In this study, we also successfully applied the proposed ViSTa-MRF method for ex-vivo scans. The obtained results from the 5-month-old and 69-year-old brain samples provided valuable insights into the myelination process during early brain development and the demyelination process in the aging brain. As part of our future work, we plan to further investigate the relationship between the estimated MWF and myelin-stained ex-vivo slabs at different cortical depths. This quantitative analysis will enable a comprehensive comparison and validation of our ViSTa-MRF-based myelin water measurements with histological myelin-stained samples.

In developing infant brains, rapid changes in relaxation times present challenges in acquiring sufficient contrast in $T_1$-weighted images for cortical segmentation and surface-based analysis (4,6,71,72). Our ViSTa-MRF technique offers an effective solution to overcome this challenge. By utilizing the quantitative $T_1$, $T_2$, and PD maps, we can synthesize $T_1$-weighted images that provide robust contrasts while significantly reducing scan time and improving motion-robustness during infant examinations, which is beneficial to generate infant substructure segmentation map. This improvement in image quality holds great promise for more accurate infant brain segmentation (73). Furthermore, the multi-contrast-weighted images synthesis eliminates the need for time-consuming structural scans in infant studies and provides an alternative when motion artifacts compromise the quality of conventional contrast-weighted images. Currently, we are utilizing quantitative maps and conventional Bloch simulations to synthesize multi-contrast weighted images, which may not fully capture magnetization transfer effect in the synthesized images (74). To address this limitation, we plan to explore the use of deep-learning based methods (75–77) for direct image synthesis from the raw k-space data in our future work, which could achieve faster and more accurate image synthesis.

## Conclusion

In this work, we have developed a 3D ViSTa-MRF technique that combines the accurate but time-consuming ViSTa technique with MR fingerprinting for whole-brain multi-parametric MRI. This approach enables us to obtain whole-brain 1mm and 660μm isotropic myelin-water fraction, quantitative $T_1$, $T_2$ and PD maps in 5 and 15 minutes, respectively. These advancements provide us with great potential to quantitatively investigate infant brain development and older adult brain degeneration.

## Acknowledgement

The authors would like to thank Dr. Vaidehi Subhash Natu, Sarah Shi Tung, Clara Maria Bacmeister and Bella Fascendini from Stanford University for their invaluable assistance in preparing the experiments. In the preparation of this manuscript, the OpenAI's Large Language Model (LLM), specifically the GPT-4 architecture, was used for grammar check. This work is supported in part by NIH research grants: R01MH116173, R01EB019437, U01EB025162, P41EB030006.

# Data and Code availability Statement

The demonstration ViSTa-MRF reconstruction scripts are available online at: https://github.com/SophieSchau/MRF_demo_ISMRM2022.
The ViSTa-MRF sequence and raw k-space datasets are available upon request.

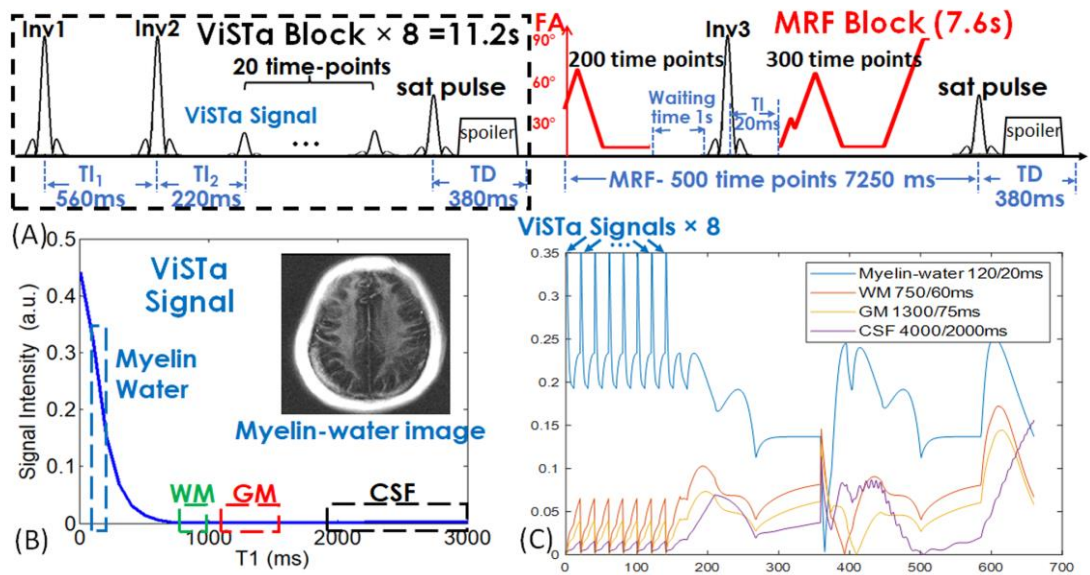

**Figure 1.** (A) Sequence diagram of 3D ViSTa-MRF. (B) Extended phase graph (EPG) simulation of the first time-point ViSTa signal. The myelin-water signal with short-$T_1$ is preserved in the ViSTa signal while the white-matter (WM), gray-matter (GM) and CSF are suppressed, which enables direct myelin-water imaging. (C) Simulated signal curves of myelin-water, WM, GM and CSF for the ViSTa-MRF sequence. FA: flip angle

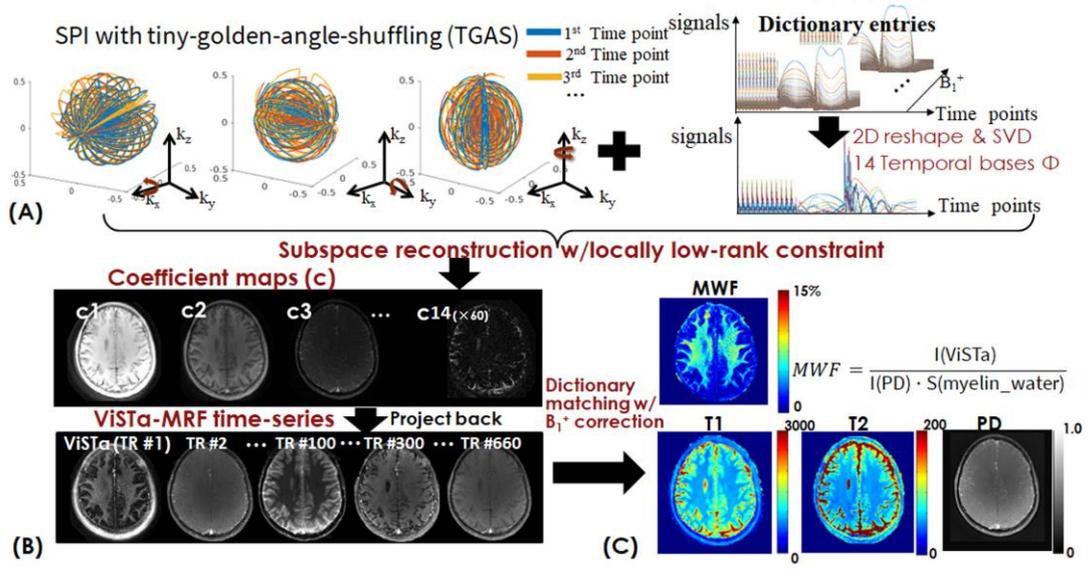

**Figure 2.** (A) Spiral-projection (SPI) sampling with tiny-golden-angle-shuffling (TGAS) trajectory and temporal components obtained from $B_1^+$ incorporated dictionary. (B) The flowchart of model-based subspace reconstruction with locally low-rank constraint. The reconstructed coefficient maps are then used to generate the ViSTa image (the first time point) and MRF time-series with voxel-by-voxel $B_1^+$ correction. (C) Template matching and $B_1^+$ correction process for $T_1$/$T_2$/PD quantification and MWF estimation.

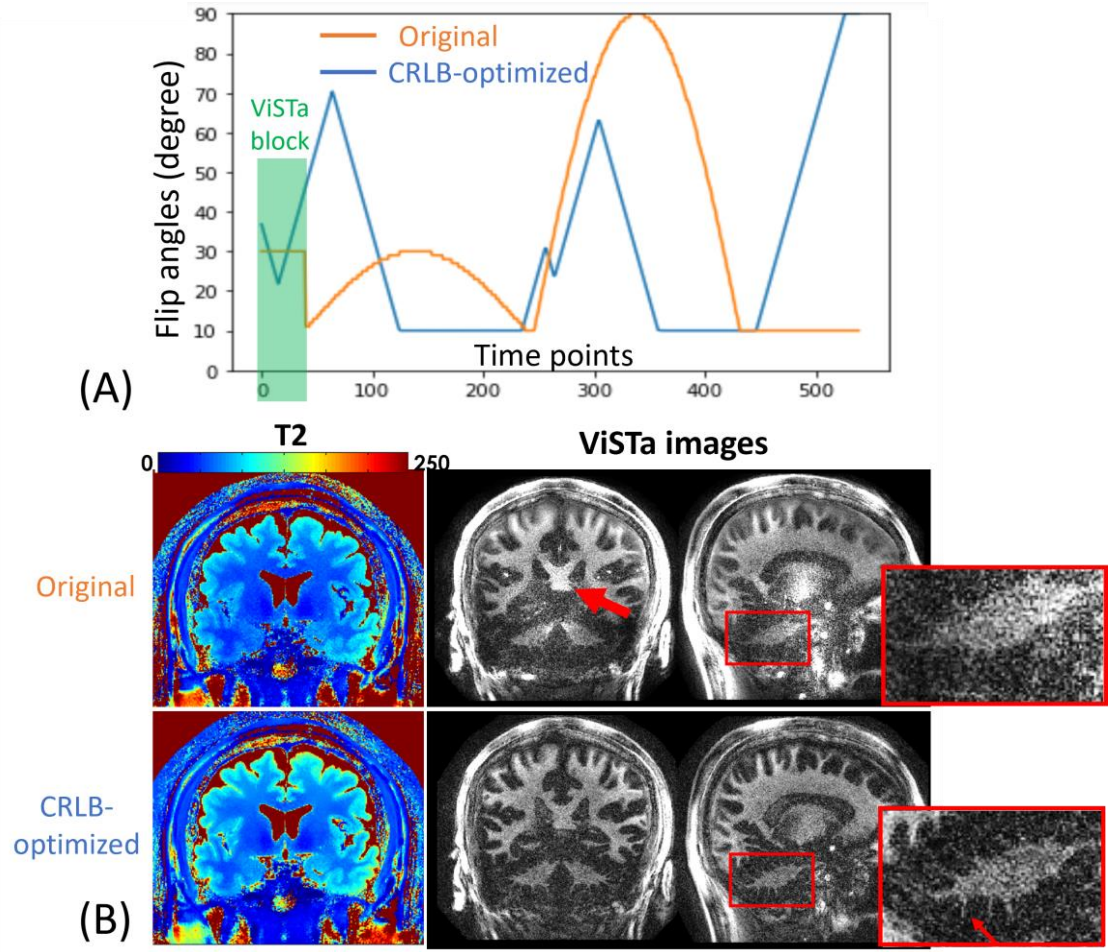

**Figure 3.** (A) Original and CRLB-optimized flip angles (FA) of the ViSTa-MRF protocol. (B) $T_2$ and ViSTa Comparisons between original ViSTa-MRF and CRLB-optimized ViSTa-MRF sequence. The zoom-in figures demonstrate that the CRLB-optimized ViSTa images exhibit higher SNR and better visualization of detailed structures in the cerebellum than the original ViSTa images.

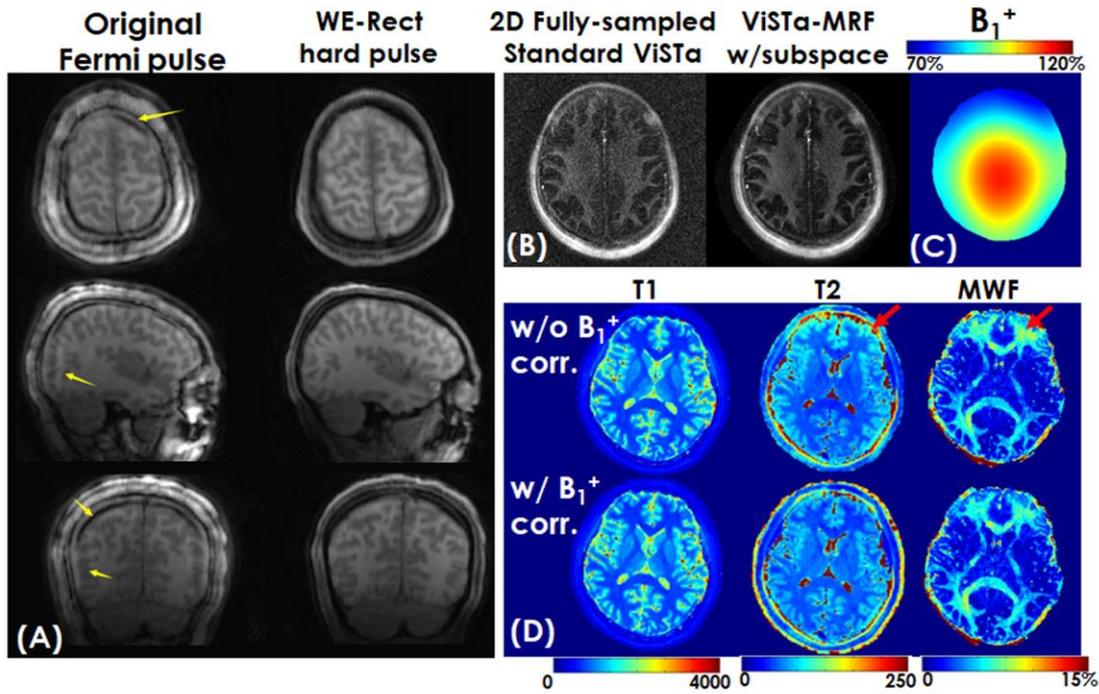

**Figure 4.** (A) Comparison of a representative MRF volume using non-selective fermi pulse and the Water-Excitation Rectangular (WE-Rect) pulse. As yellow arrows indicate, the fat artifacts are mitigated using the WE-Rect pulse. (B) The comparison between a fully sampled standard 2D ViSTa sequence and ViSTa-MRF with subspace reconstruction. (C) Pre-scanned and spatial-smoothed $B_1^+$ map and (D) reconstructed $T_1$, $T_2$, and MWF without (first row) and with (second row) $B_1^+$ correction. With $B_1^+$ correction, the estimated $T_2$ and MWF maps in regions indicated by the red arrows are more uniform compared to the results without $B_1^+$ correction.

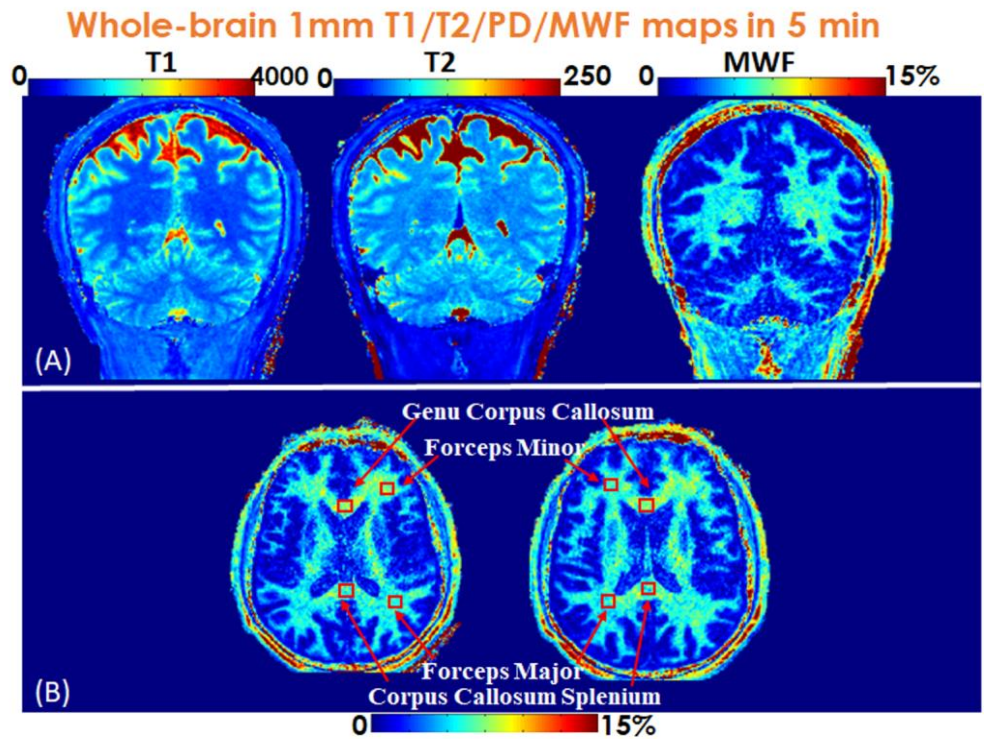

**Figure 5.** (A) Whole-brain 1-mm iso $T_1$, $T_2$ and MWF maps in coronal views. (B) Two representative slices of MWF maps in axial views.

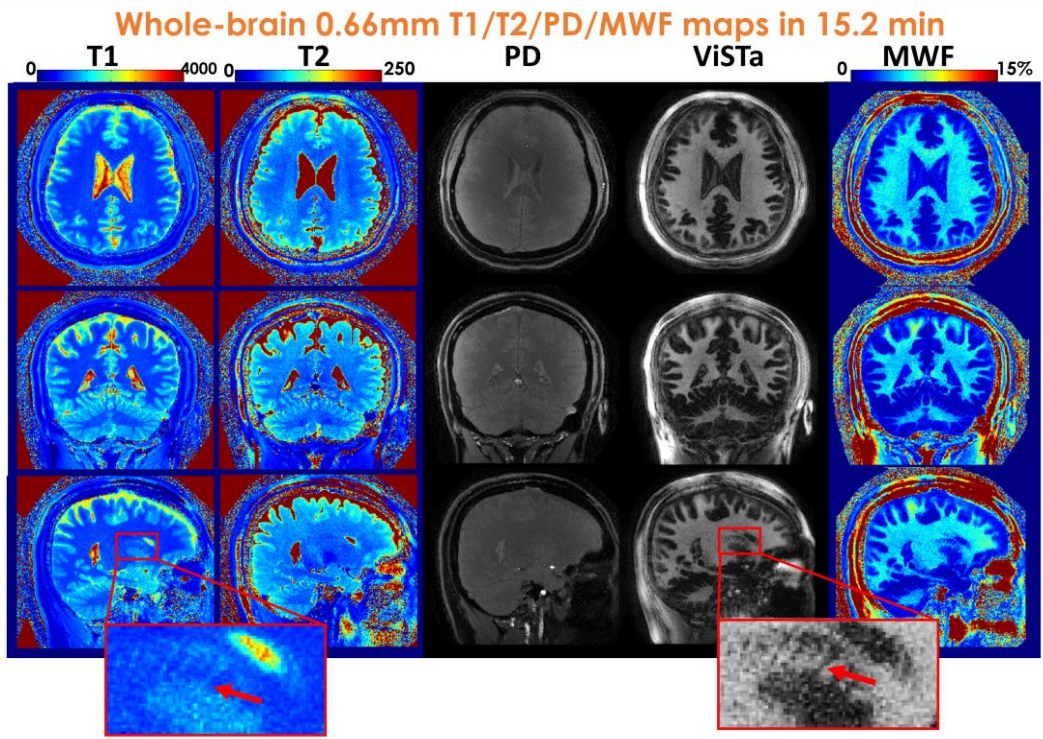

**Figure 6.** Whole-brain 0.66mm-iso $T_1/T_2$/PD/ViSTa and MWF maps in three orthogonal views.

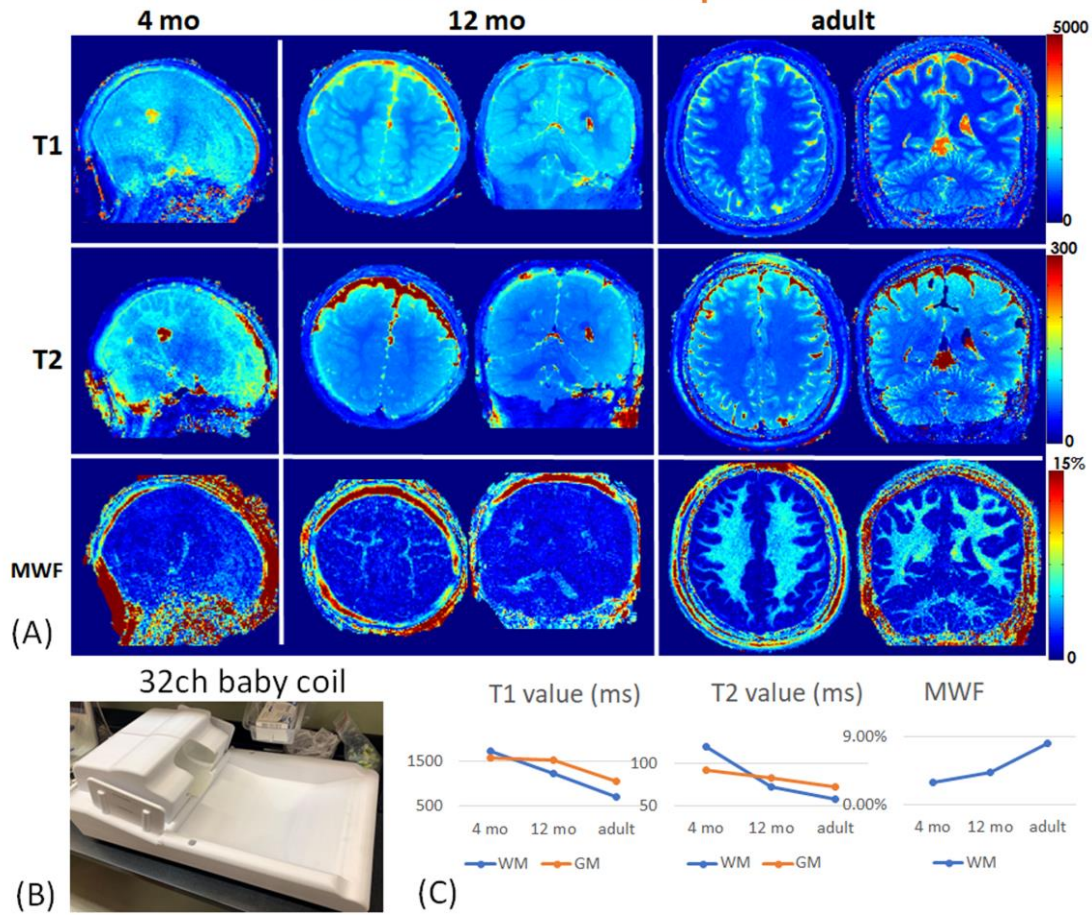

**Figure 7.** (A) Whole-brain 1.0 mm isotropic $T_1$, $T_2$ and MWF maps of a 4-month infant, a 12-month infant and an adult as a reference. (B) The baby data were acquired with a custom 32-channel tight-fitting baby coil. (C) Both $T_1$ and $T_2$ values decrease while the MWF increases with brain development, indicating the brain dynamic process of myelination.

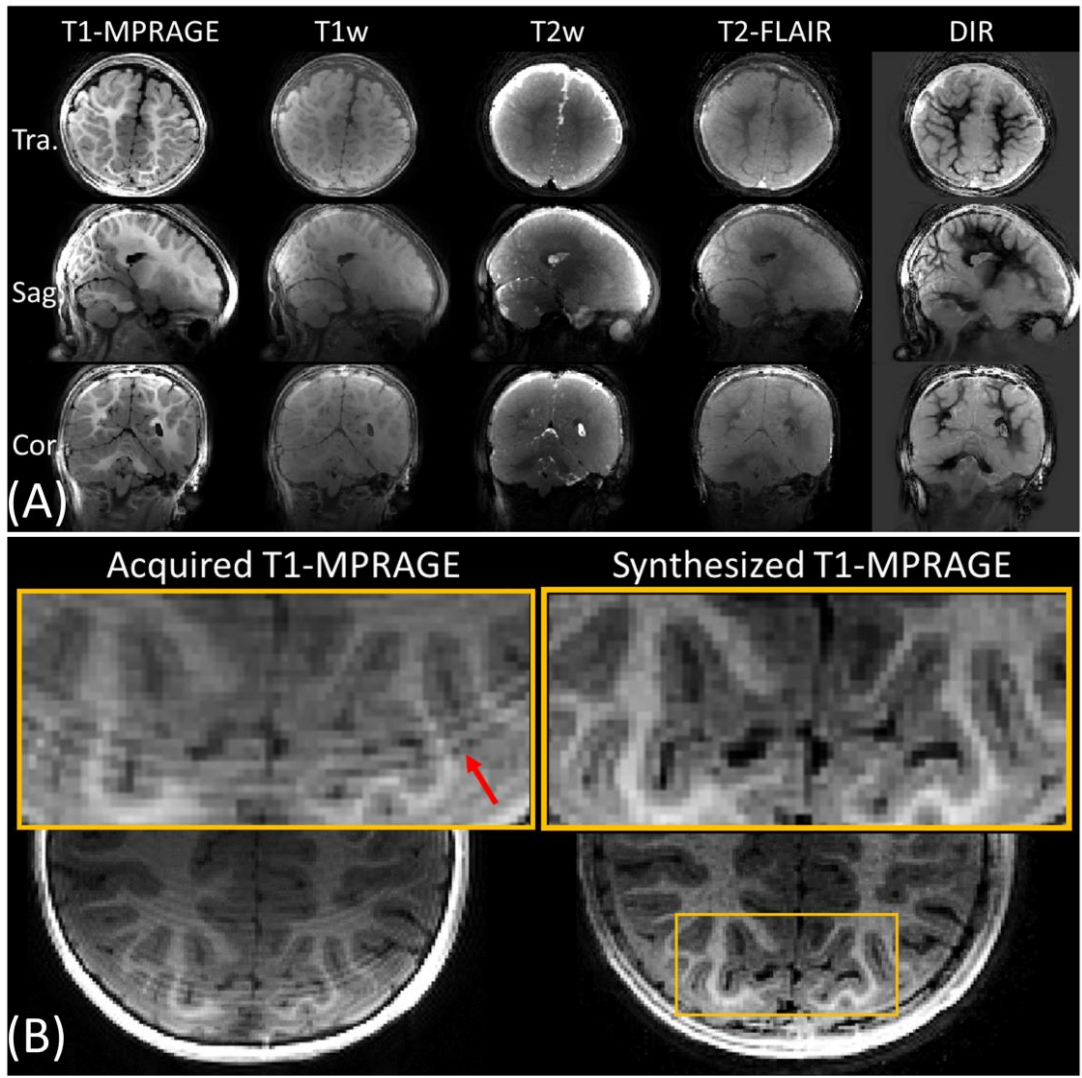

**Figure 8.** (A) Synthesized whole-brain $T_1$-MPRAGE, $T_1$-& $T_2$-weighted, $T_2$-FLAIR and double-inversion-recovery (DIR) images using the 5-minute 1mm-iso ViSTa-MRF protocol from a 12-month-old infant data. (B) Comparison between an acquired $T_1$-weighted MPRAGE image (left) and a ViSTa-MRF synthesized $T_1$-weighted image (right).

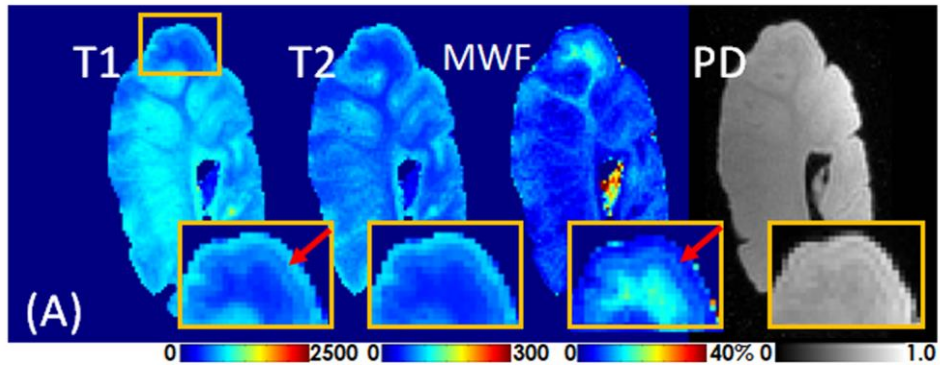
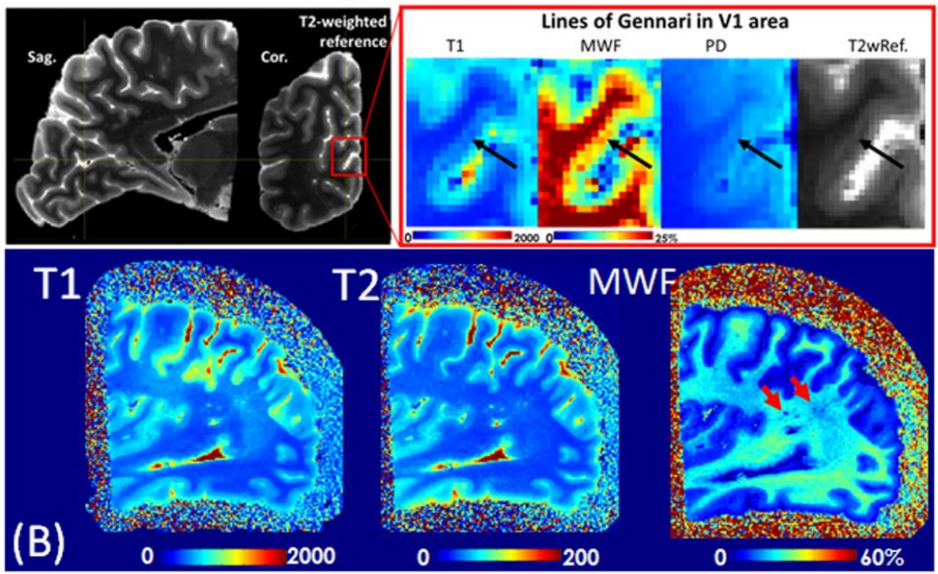

**Figure 9.** 0.50-mm iso $T_1/T_2$/PD and MWF maps of (A) a coronal slab obtained from a 5-month-old post-mortem brain and (B) a left occipital lobe obtained from a 69-year-old post-mortem brain. The total acquisition time is 57 minutes. $T_1/T_2$/MWF/PD maps shows (A) the myelinated lines in cortical layers and (B) lines of Gennari in V1 region with decreased $T_1$ & PD and increased MWF values (indicated by black arrows) and the "dark dots" in MWF and increased $T_1$ and $T_2$ values (indicated by red arrows).

# Supporting Information

**Table S1.** MWF comparison between the proposed ViSTa-MRF method and the literature values in splenium, forceps Major/minor, and genu corpus callosum regions.

|  | Literature values (Ref. (25)) | ViSTa-MRF MWF |
|---|---|---|
| Splenium | 5.4%±0.3% | 5.8%±1.2% |
| Genu | 5.8%±0.4% | 5.7%±0.8% |
| Forceps Major | 5.2%±0.4% | 5.2%±1.0% |
| Forceps Minor | 5.0%±0.4% | 5.3%±1.0% |

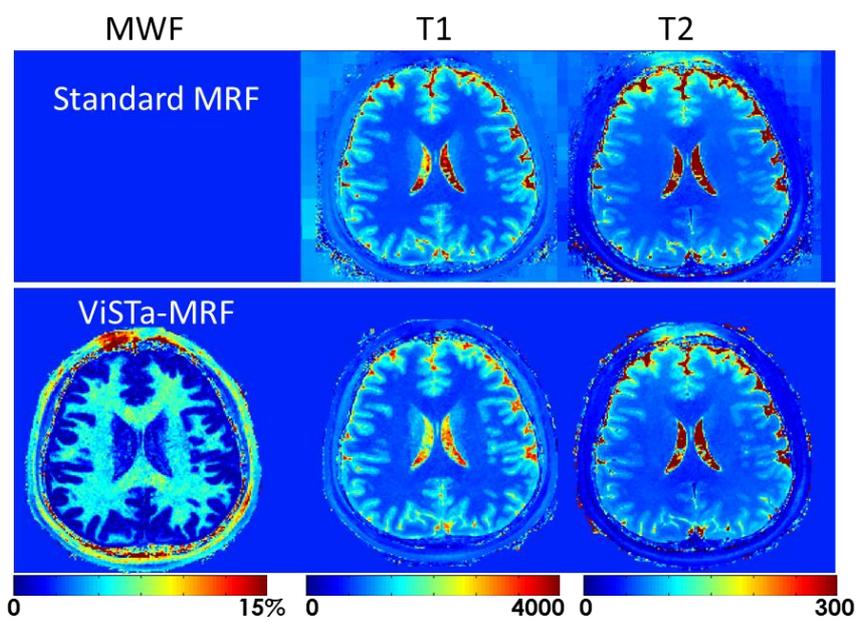

**Figure S1.** $T_1$ and $T_2$ comparison between 1mm-iso ViSTa-MRF and standard MRF.

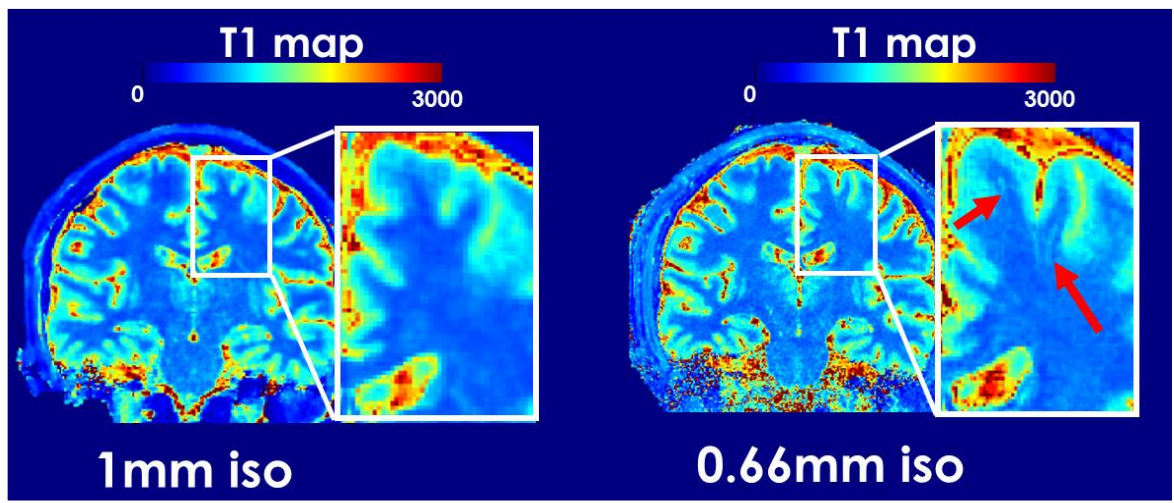

**Figure S2.** $T_1$ comparison between 1mm and 0.66-mm data. The higher resolution in the 0.66-mm dataset can aid in better visualization of subtle brain structures such as small sulci and the periventricular space as indicated by the red arrows.

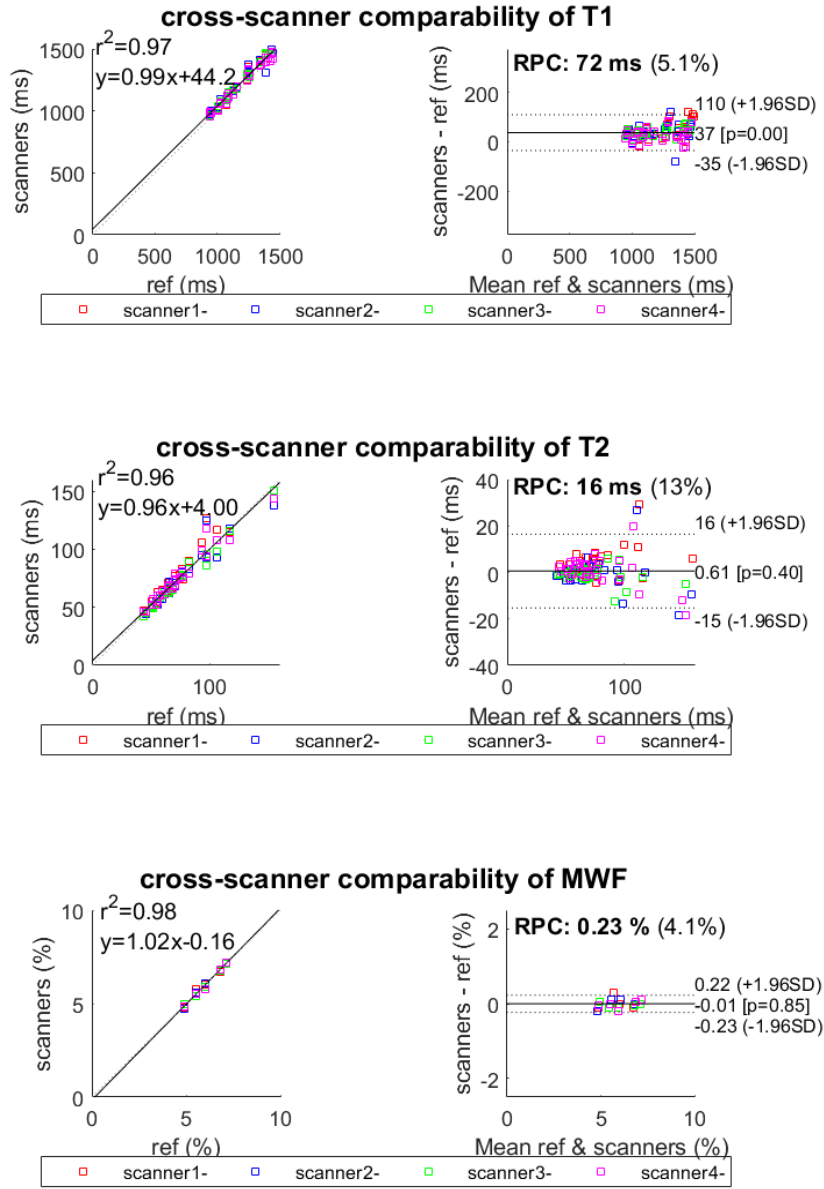

**Figure S3.** Cross-scanner comparability of ViSTa-MRF results.